\documentclass[aps,twocolumn,showpacs]{revtex4}
\usepackage{graphicx}
\usepackage{amsfonts}

\newcommand{\fft}[2]{{\frac{#1}{#2}}}

\begin{document}

\preprint{hep-th/0610147}

\title{Spin chain from marginally deformed $AdS_3\times S^3$}

\author{Wen-Yu Wen} \email{steve.wen@gmail.com}
\affiliation{Department of Physics and Center for Theoretical
Sciences, National Taiwan University, Taipei 106, Taiwan}


\begin{abstract}
We derive a spin chain Hamiltonian from a fast spinning string in
the marginally deformed $AdS_3\times S^3$.  This corresponds to a
closed trajectory swept out by the $SU(2)$ or $SL(2)$ spin vector
on the surface of one-parameter deformed two-sphere or hyperboloid
in the background of anisotropic magnetic field interaction. In
the limit of small deformation, a class of general Landau-Lifshitz
equation with a nontrivial anisotropic matrix can be derived.
\end{abstract}


\maketitle


\section{Introduction}

The anti-de Sitter/conformal field theory (AdS/CFT) correspondence
has revealed deep relation between string theory and gauge theory,
in particular the correspondence between IIB strings on
$AdS_5\times S^5$ and ${\cal N}=4$ super Yang-Mills theory (SYM)
at the 't Hooft's large N
limit\cite{Maldacena:1997re,Gubser:1998bc,Witten:1998qj}. While
the correspondence was mostly tested for those
Bogomol'nyi-Prasad-Sommerfield (BPS) states where partial
supersymmetry protects it against quantum
correction\cite{Aharony:1999ti}, semiclassical analysis on the
near-BPS sector is also considered in
Ref.\cite{Berenstein:2002jq}; the integrability of this
Berenstein-Madalcena-Nastase (BMN) limit allows for quantitative
tests of correspondence beyond BPS states, where the energy of
classical string solutions is compared to the anomalous dimension
of SYM operators with large {\it R}-charge.  On the other hand,
intimate relation between SYM dynamics and integrable spin chain
was realized in Ref.\cite{Beisert:2003ea}, because the planar
limit of the dilatation operator was identified with the
Hamiltonian of integrable spin chains.

The string/spin chain correspondence was pushed further in
Ref.\cite{Kruczenski:2003gt}, where a classical string spinning on
$S^3\in S^5$ were identified with semiclassical coherent states in
the $SU(2)$ spin chain system. Later this identification was
explored in the full $SU(3)$ sector\cite{Hernandez:2004uw} and
$SL(2)$ sector\cite{Bellucci:2004qr}.  A few examples of
generalization were discussed in the past: the fast spinning
string in the marginally $\beta$-deformed ${\cal
N}=4$\cite{Lunin:2005jy} corresponds to an anisotropic XXZ spin
chain\cite{Frolov:2005ty}. The Melvin's magnetic-deformed
background was also studied in Ref.\cite{Huang:2006bh}.

Here we are interested in another integrable system, $AdS_3\times
S^3$ background\cite{Gomis:2002qi,Lunin:2002fw}.  As discussed in
Ref.\cite{Israel:2004vv}, asymmetric marginal deformations of
$G=SU(2)_k$ and $SL(2,R)_{k}$ Wess-Zumino-Witten (WZW) models
create new families of exact string vacua. The corresponding
geometry is a deformed $S^3$ or $AdS_3$ which includes, in
particular, geometric coset $S^2$ or $AdS_2$. In general, the
effect of deformation is to turn on an electromagnetic field along
some Cartan direction inside $G$, which in turn induces a
gravitational back-reaction on the metric and the three-form
antisymmetric tensor.

The plan of this report is as follows.  We derive spin chain
models in the fast spinning limit from the deformed $SU(2)_k$ in
section II, and deformed $SL(2)_k$ in section III.  They
correspond to the deformed configuration space of spin vector, a
squashed two-sphere and hyperboloid, respectively.  In section IV,
we take the limit of small deformation and find the system can be
exactly derived from a class of general Landau-Lifshitz equation
with a nontrivial anisotropic matrix. In section V, we have a
discussion and comments.

\section{Deformed $SU(2)_k$ spin chain model}

\begin{figure}\label{fig_1}
\includegraphics{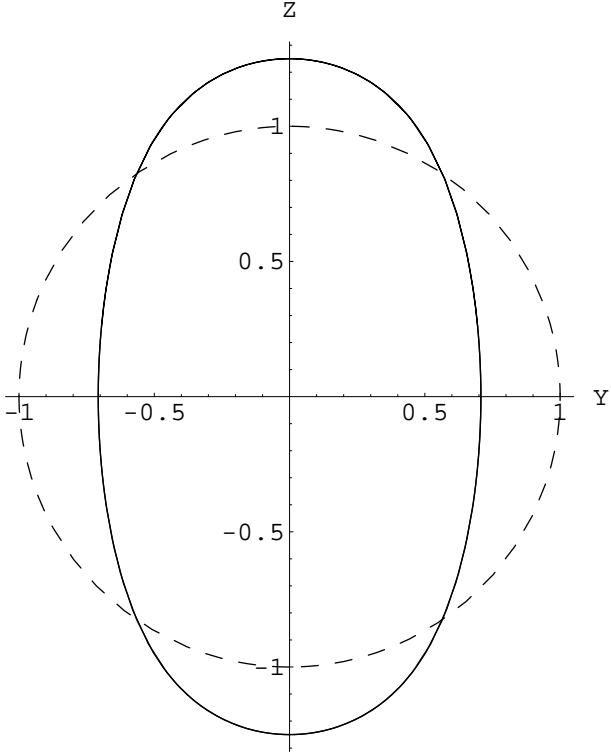}
\caption{Configuration space of spin vector (solid line) is
deformed from unit two-sphere(dashed line). Here only projection
on YZ-plane is displayed.}
\end{figure}

The $SU(2)$ manifold is a three-sphere of unit radius.  The
$SU(2)_k$ WZW model then corresponds to a three-sphere of radius
$k$ at classical level.  In Ref.\cite{Israel:2004vv}, the authors
discussed asymmetric marginal deformation in the background of
three-sphere which breaks the bosonic affine algebra
$SU(2)_L\times SU(2)_R$ into $U(1)_L\times SU(2)_R$.  The deformed
metric is given by
\begin{equation}\label{metric_deform}
ds^2=\fft{k}{4}[-dt^2+d\beta^2+\sin^2{\beta}d\alpha^2+(1-2H^2)(d\gamma+\cos{\beta}d\alpha)^2]
\end{equation}
We request $0\le H^2 \le 1/2$ to avoid non-unitary gauge field
($H^2<0$) and closed time-like geodesics ($H^2>1/2$).  For trivial
$H=0$ and at level $k=1$, one recovers the unit round $S^3$ as
Hopf fibration.  For maximal $H^2=1/2$, the $S^1$ fiber
degenerates and we are left with $S^2$.

Now we consider a spinning string by sending $\alpha \to t +
\alpha$ in Eq.(\ref{metric_deform}), with gauge choice
$t=\kappa\tau$, the Polyakov action reads,
\begin{eqnarray}\label{action}
S =&& \fft{\sqrt{\lambda}k}{16\pi}\int\int{d\tau d\sigma}
-2H^2\cos^2{\beta}\kappa^2\nonumber\\
&&-(1-2H^2\cos^2{\beta})\alpha'^2-\beta'^2- (1-2H^2)\gamma'^2\nonumber\\
&&-2(1-2H^2)\cos{\beta}\alpha'\gamma'+2(1-2H^2\cos^2{\beta})\kappa
\dot{\alpha}\nonumber\\
&&+2(1-2H^2)\cos{\beta}\kappa
\dot{\gamma}+(1-2H^2\cos^2{\beta})\dot{\alpha}^2\nonumber\\
&&+\dot{\beta}^2+ (1-2H^2)\dot{\gamma}^2
+2(1-2H^2)\cos{\beta}\dot{\alpha}\dot{\gamma},
\end{eqnarray}
where the dot and prime are derivative with respect to worldsheet
time and space coordinates.  Vanishing of diagonal components of
the stress-energy tensor puts on a Virasora constraint:
\begin{eqnarray}
&&(1-2H^2\cos^2{\beta})\dot{\alpha}\alpha'+\dot{\beta}\beta'\nonumber\\
&&+(1-2H^2)\dot{\gamma}\gamma'+2(1-2H^2\cos^2{\beta})\kappa\alpha'\nonumber\\
&&+2(1-2H^2)\cos{\beta}(\kappa\gamma'+\dot{\alpha}\gamma'+\alpha'\dot{\gamma})=0.
\end{eqnarray}
Now we take fast spinning limit $\kappa \to \infty,
\dot{X^{\mu}}\to 0$ but keep $\kappa \dot{X^{\mu}}$ finite for
$\mu \neq t$.  At this limit, the contribution of kinetic term,
i.e. $\dot{X}^2$, is ignorable thus can be dropped.  In the other
words, the string is {\it frozen} and energy is contributed mostly
from curving the string.  The the relevant part of
Eq.(\ref{action}) becomes
\begin{eqnarray}\label{action_1}
S&&=\fft{\sqrt{\lambda}k}{16\pi}\int\int d\tau d\sigma
-2H^2\cos^2{\beta}\kappa^2+\pi_{\alpha}\dot{\alpha}\nonumber\\
&&+\pi_{\gamma}\dot{\gamma}-\beta'^2-\Delta
\gamma'^2,\nonumber\\
&&\Pi_{\alpha}\equiv \int{d\sigma}\pi_{\alpha}= 2\kappa\int{d\sigma} (1-2H^2\cos^2{\beta}), \nonumber\\
&&\Pi_{\gamma}\equiv
\int{d\sigma}\pi_{\gamma}=2(1-2H^2)\kappa\int{d\sigma}
\cos{\beta},\nonumber\\
&&\Delta\equiv\fft{(1-2H^2)\sin^2{\beta}}{1-2H^2\cos^2{\beta}},
\end{eqnarray}
where the Virasora constraint has been applied to eliminate
$\alpha'$.  Two comments are in order: At first, we notice that
there is no explicit time derivative in the Hamiltonian.
Therefore, one may find a mapping $\Gamma: \sigma \to \tilde{S}^2$
from string worldsheet coordinate $\sigma$ to the deformed
two-sphere.  The periodicity of closed string ensures this
$\Gamma(\sigma)$ a closed trajectory on the sphere surface.  To
see this explicitly, one may introduce the spin vector pointing to
an arbitrary point on the surface,
\begin{equation}
\vec{n}=(\sqrt{\Delta}\cos{\gamma},\sqrt{\Delta}\sin{\gamma},n_z(\beta,H))
\end{equation}
such that the following constraint is satisfied
\begin{equation}\label{constraint_1}
((\sqrt{\Delta})')^2+(n'_z)^2=1.
\end{equation}
In generic, there may not be an analytic solution for
$n_z(\beta,H)$ but it can be solved numerically. It is not
surprising that the norm $|\vec{n}|$ is not a constant as in
\cite{Kruczenski:2003gt} but a function of $\beta$ and $H$ due to
the deformation.  However, the constraint (\ref{constraint_1})
implies $\Gamma$ is an uniform map.  This reflects the very fact
that string has no infrastructure for stretching.  Figure 1 shows
that the configuration space of the spin vector is deformed from a
unit two-sphere.

Second, the conjugate variables $\Pi$'s are angular momenta
associated with $\alpha$ and $\gamma$.  Those terms can be further
removed by the other Virasora constraint due to tracelessness of
stress-energy tensor.  However, to keep them in the above action
is useful in comparison with spin chain model. The angular
momentum $\Pi_{\alpha}$ gives us information of length of spin
chain,
\begin{equation}
J=J_0-\int{d\tilde{\sigma} 2H^2\cos^2{\beta}},
\end{equation}
where the total length of chain for vanishing $H$ is defined as
\begin{equation}
J_0=\fft{\sqrt{\lambda}k}{16\pi}2\kappa\int{d\sigma}\equiv
\int{d\tilde{\sigma}}.
\end{equation}
The term $\Pi_{\gamma}\dot{\gamma}$ is called the Wess-Zumino term
and it is related to the surface area bounded by $\Gamma$ in the
undeformed case.  Without knowing the explicit form of $n_z$, one
can still write the spin vector with small $H$ expansion,
\begin{equation}
n_z\simeq \cos{\beta}+\cos^3{\beta}H^2+{\cal O}(H^4).
\end{equation}
The Hamiltonian from (\ref{action_1}) becomes
\begin{equation}\label{hamilton}
{\cal H} =\int\int{dt d\tilde{\sigma}} H^2\cos^2{\beta}
+\fft{\lambda k^2}{128\pi^2}\int\int{dt d\tilde{\sigma}}
\beta'^2+\Delta \gamma'^2,
\end{equation}
where the derivative of $\beta$ and $\gamma$ is with respective to
$\tilde{\sigma}$.  Up to a rescale of coordinates, it is easy to
see that the above agrees with Ref.\cite{Kruczenski:2003gt} for
vanishing $H$.  The first term in Eq.(\ref{hamilton}) shows a
quadrupole type interaction
\begin{equation}
{\cal H}_{int} \simeq  J_0\int\int (\vec{n}\cdot\vec{B})^2.
\end{equation}
Here we have claimed that the deformation of $SU(2)_k$ manifold
can be described by an external magnetic field $B_z=H$ with the
quadrupole type interaction\footnote{It is interesting to compare
this with interaction obtained in Ref.\cite{Huang:2006bh}, where
the dipole type interaction also appears.}. At last, we would like
to comment on the NS-NS B-field which has been ignored before. The
deformed {\it B}-field can be obtained locally\cite{Israel:2004vv}
with
\begin{equation}
B_{[2]}=\frac{k}{4}(1-2H^2)\cos{\beta}d\alpha\wedge d\gamma,
\end{equation}
which couples to string worldsheet in the way
\begin{equation}
S_B=\frac{\sqrt{\lambda}}{4\pi}\int\int{d\tau d\sigma}
\epsilon^{ab}\partial_aX^{\mu}\partial_bX^{\nu}B_{\mu\nu}.
\end{equation}
Since we have taken the limit where $\dot{X^{\mu}}\to  0$, it is
obvious $S_B$ becomes irrelevant in this limit and action
(\ref{action_1}) is still valid.

\section{Deformed $SL(2)_k$ spin chain model}

\begin{figure}\label{fig_2}
\includegraphics{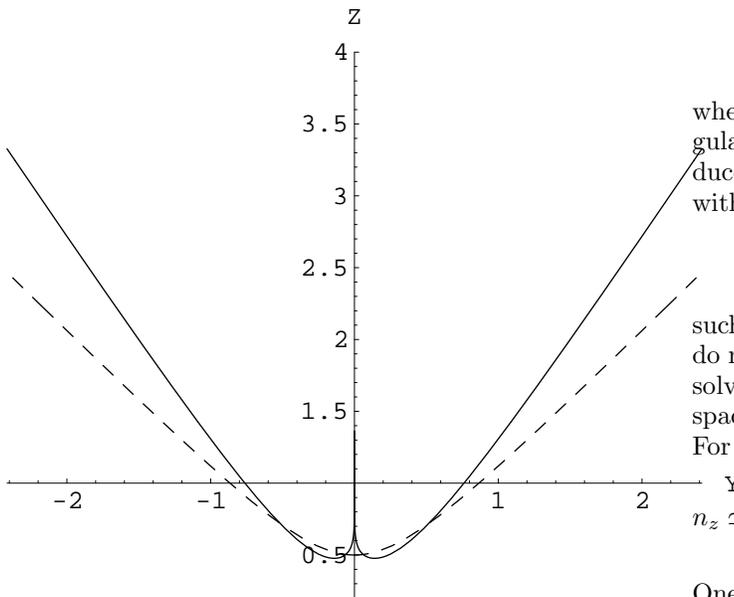}
\caption{Configuration space of spin vector (solid line) is
deformed from hyperboloid (dashed line). Here only projection on
YZ-plane is displayed.}
\end{figure}

In this section, we consider the hyperbolic deformation of
$AdS_3$, or correspondingly $SL(2)_k$ WZW
model\cite{Israel:2004vv}. The isometry after deformation reduces
to $SO(1,1)_L\times SL(2)_R$. The resulting metric reads,
\begin{equation}
ds^2 = \frac{k}4 \left[ d\rho^2 - \cosh^2 \rho dt^2 + (1 - 2 H^2)
(d\phi + \sinh \rho dt)^2 +d\gamma^2 \right],
\end{equation}
where the string also sits at fixed $\alpha$ and $\beta$ inside
$S^3$. Now we consider fast spinning string, say, let $\gamma \to
t+ \gamma$, and at the same time rescale $d\phi\to
\sinh{\rho}d\phi$. With gauge choice $t=\kappa \tau$ and taking
the limit $\kappa\to \infty, \dot{X}^{\mu}\to 0$ but keeping
$\kappa \dot{X}^{\mu}$ finite, one reaches
\begin{eqnarray}\label{action_2}
&&S=\fft{\sqrt{\lambda}k}{16\pi}\int\int d\tau d\sigma
-2H^2\sinh^2{\rho}\kappa^2 \nonumber\\
&&+2(1-2H^2)\sinh^2{\rho}\kappa\dot{\phi}+2\kappa\dot{\gamma}-\rho'^2-\Delta_h
\phi'^2,\nonumber\\
&&\Delta_h\equiv(1-2H^2)\sinh^2{\rho}[1+(1-2H^2)\sinh^2{\rho}],
\end{eqnarray}
where the term associated with $\dot{\gamma}$ gives us the same
angular momentum as the undeformed one.  One may introduce the
spin vector moving over a deformed hyperboloid with signature
$(-,-,+)$.
\begin{equation}
\vec{n}=(\sqrt{\Delta_h}\cos{\phi},\sqrt{\Delta_h}\sin{\phi},n_z(\rho,H))
\end{equation}
such that $-((\sqrt{\Delta_h})')^2+(n'_z)^2=-1$ is satisfied. We
again do not know the analytic form for $n_z(\rho,H)$ but it can
be solved numerically.  Figure 2 shows that the configuration
space of the spin vector is deformed from a hyperboloid.  For
small $H$ expansion, we obtain
\begin{equation}
n_z\simeq
\fft12\cosh{2\rho}+[\fft12(\log{\coth{\rho}}-2\cosh{2\rho})-
\fft{1}{4 \cosh{\rho}}]H^2+{\cal O}(H^4).
\end{equation}
One can also read the Hamiltonian from Eq.(\ref{action_2}),
\begin{equation}
{\cal H} =\int\int{dt d\tilde{\sigma}} H^2\sinh^2{\rho}
+\fft{\lambda k^2}{128\pi^2}\int\int{dt
d\tilde{\sigma}}\rho'^2+\Delta_h \phi'^2,
\end{equation}
where the first term can be rewritten as
\begin{equation}
{\cal H}_{int} \simeq  J_0\int\int H^2 (n_z-\fft12).
\end{equation}

\section{Landau-Lifshitz equation as $H\to 0$ limit}
In the previous discussion, we have absorbed the infinite $\kappa$
by redefinition of $\sigma$ on the string worldsheet such that the
Hamiltonian density appears finite.  String sees the deformation
via both interaction term ${\cal H}_{int}$ and $\Delta_{(h)}$ in
front of $\gamma'^2$ or $\phi'^2$.  In order to go around the
complication due to deformation, one may take a new double scaling
limit where both $H\to 0$ and $\kappa \to \infty$ limits are
taken, but keep $\kappa H$ and $\kappa \dot{X^{\mu}}$ finite. In
this new limit, one is able to keep nontrivial interaction term
but recover the round sphere as in Ref.\cite{Kruczenski:2003gt}
due to $\Delta\to \sin^2{\beta}$(or hyperboloid as $\Delta_h\to
(\fft12\sinh{2\rho})^2$). It is convenient to rescale $\kappa H
\to H$ and $\kappa\dot{X}\to\dot{X}$ and we will only focus on the
deformed $SU(2)_k$ in the following discussion. The equations of
motion for $\beta$ and $\gamma$ after rescaling are given by
\begin{eqnarray}\label{eom}
&&\beta'' -
\sin{\beta}\dot{\gamma}-\sin{\beta}\cos{\beta}(\gamma')^2+H^2\sin{2\beta}=0\nonumber\\
&&\sin{\beta}\dot{\beta}+(\sin^2{\beta}\gamma')'=0
\end{eqnarray}
It has been long established that the classical Heisenberg spin
chain is completely integrable and its equation of motion is a
particular case of a more general Landau-Lifshitz
equation\cite{Kazakov:2004qf}
\begin{equation}\label{ll}
\partial_t \vec{S} = \vec{S} \times \partial^2_{\sigma}\vec{S} + \vec{S}\times
{\cal J}\vec{S},
\end{equation}
where
\begin{equation}\label{spin_vec}
\vec{S}=(\sin{\beta}\cos{\gamma},\sin{\beta}\sin{\gamma},\cos{\beta}),
\end{equation}
${\cal J}$ is the unit matrix for the isotropic
case\cite{Kruczenski:2003gt}, but expected to be nontrivial for
the anisotropic one.  Indeed, a straight calculation shows that
(\ref{eom}) can be derived from (\ref{ll}), provided
(\ref{spin_vec}) and the anisotropic matrix ${\cal J}=diag[
j_1,j_2,j_3]$ where
\begin{equation}
j_1=j_2=1,\qquad j_3=1-2H^2.
\end{equation}
This general Landau-Lifshitz equation can be seen as continuous
limit of the inhomogeneous Heisenberg spin chain model
\begin{equation}\label{spinchain}
{\cal H}=-J_0\sum_i{S_i\cdot S_{i+1}}+ \sum_i{ a S^3_{i}\cdot
S^3_{i+1}},
\end{equation}
where the anisotropy parameters $a \propto j_3-j_1=j_3-j_2=-2H^2$.
Integrability of (\ref{ll}) provides the evidence that this
marginally deformed $SU(2)_k$ background is integrable, at least
in the double scaling limit of fast spinning and small
deformation.

\section{Discussion}
In this report, we have investigated the fast spinning limit of
classical string in the deformed background $AdS_3\times S^3$. In
particular, we have considered the asymmetric marginally deformed
$SU(2)_k$ and $SL(2)_k$ WZW model.  Four comments are in order.
Since only the component $j_3$ is deviated from identity in the
case of deformed $SU(2)_k$, it is curious if there exists a more
general correspondence between nontrivial $j_i$'s and other kinds
of deformation.  However, if the deformed $SU(2)_k$ discussed in
Ref.\cite{Israel:2004vv} is unique, those other deformation, if
exist, must be relevant or irrelevant.  In that sense,
integrability makes the system still under control even outside
the reach of marginal deformation. Second, the $H\to 0$ limit
implies that we may use the same coherent states as constructed in
Refs.\cite{Kruczenski:2003gt,Bellucci:2004qr,Tseytlin:2004xa} to
build up a discrete sigma model, whose continuous limit will
recover the action of fast spinning string.  Thirdly, if one could
start with a deformed WZW sigma model\cite{Israel:2004vv} whose
target space is given by (\ref{metric_deform}) and calculate the
dilatation operator, the same spinning string can be constructed
whose thermodynamic limit may give rise to the general
Landau-Lifshitz equation (\ref{ll}). At last, solutions to
Eq.(\ref{ll}) has been intensely studied in recent decades.  In
particular, it has been found that the (multi-)soliton solutions
may correspond to (multi-)magnon-like excitation first discussed
in Ref.\cite{Hofman:2006xt}.  We plan to come back to these issues
in future publications\cite{Wen:2007}.

\section*{ACKNOWLEDGMENTS} I am grateful to W.~H.~Huang for
clarification on previous work on the subject and inspiration for
this project. I also would like to thank C.~H.~Chen, P.~M.~Ho,
S.~Teraguchi, D.~Tomino, C.~H.~Yeh and S.~Zeze for useful
discussion and comments.  The author is supported in part by the
Taiwan's National Science Council under Grant No.
NSC95-2811-M-002-013.

\end{document}